\documentclass[twocolumn]{aastex631}

\usepackage{CJKutf8}
\usepackage{color}

\usepackage{graphicx}	
\usepackage{amsmath}	
\usepackage{amssymb}	

\newcommand\lsim{\mathrel{\rlap{\lower4pt\hbox{\hskip1pt$\sim$}}
\raise1pt\hbox{$<$}}}
\newcommand\gsim{\mathrel{\rlap{\lower4pt\hbox{\hskip1pt$\sim$}}
\raise1pt\hbox{$>$}}}

\shortauthors{Faridani et al.}

\graphicspath{{./}{figures/}}

\shorttitle{Let's Sweep!}
\shortauthors{Faridani et al.}
\graphicspath{{./}{figures/}}

\begin{document}

\title[Let's Sweep]{Let's Sweep: The Effect of Evolving $J_2$ on the Resonant Structure of a Three-Planet System}

\author[0000-0003-3799-3635]{Thea H. Faridani}
\correspondingauthor{Thea H. Faridani}
\email{thfaridani@astro.ucla.edu}
\affiliation{Department of Physics and Astronomy, University of California, Los Angeles, CA 90095, USA}
\affiliation{Mani L. Bhaumik Institute for Theoretical Physics, Department of Physics and Astronomy, UCLA, Los Angeles, CA 90095, USA}

\author[0000-0002-9802-9279]{Smadar Naoz}
\affiliation{Department of Physics and Astronomy, University of California, Los Angeles, CA 90095, USA}
\affiliation{Mani L. Bhaumik Institute for Theoretical Physics, Department of Physics and Astronomy, UCLA, Los Angeles, CA 90095, USA}



\author[0000-0001-8308-0808]{Gongjie Li}
\affiliation{School of Physics, Georgia Institute of Technology, Atlanta, GA 30332, USA}
\author[0009-0002-3550-2310]{Nicholas Inzunza}
\affiliation{Department of Physics and Astronomy, University of California, Los Angeles, CA 90095, USA}
\affiliation{Mani L. Bhaumik Institute for Theoretical Physics, Department of Physics and Astronomy, UCLA, Los Angeles, CA 90095, USA}



\begin{abstract}

Short and ultra-short planets are a peculiar type of exoplanets with periods as short as a few days or less. Although it is challenging to detect them, already several are observed with many additional candidates. If these planets have formation pathways to their longer period counterparts, they are predicted to reside in multi-planet systems. Thus, gravitational perturbation from potential planetary neighbors may affect their orbital configuration. However, due to their close proximity to their host star, they are also subjects to general relativity precession and torques from the stellar spin quadrupole moment ($J_2$).  Here we show that an evolving $J_2$ due to magnetic braking, affects the magnitude and location of secular resonances of the short period planet in a multi-planet system. Thus, driving the short period planet into and out of a secular resonance, exciting the planet's eccentricity and inclination. The  high inclination can hinder transit observation, and, in some cases, the high eccentricity may result in an unstable configuration. We propose that evolving $J_2$ in a multi-planet system can be critical in understanding the detectability and stability of short-period planets. 
\end{abstract}



\section{Introduction}

It has been shown over the past two decades that systems of tightly packed period planets, inwards to the orbit of Mercury, are common in our galaxy \citep[e.g.,][]{Howard+10,Howard+12,Lissauer+11,Winn+15}. Whether accompanied by far-orbiting companion planets or not, these systems are home to myriad interesting dynamics from planet-planet interactions to star-planet interactions  \citep[e.g., ][]{Faridani+22, wei+21,Fabrycky+07,Naoz+13,Liu+15,Mardling+04,Plavchan+15,Spalding+16,Becker+17}. 

Among these systems exists a population of Ultra-Short Period planets (USPs) which have orbits lasting $24$ hours or fewer. Although these planets are intrinsically rare  \citep[orbiting only about one star in two hundred][]{Sanchis-Ojeda+14}, they are significantly easier to detect than planets with similar mass/size at longer orbital periods because of their frequent transits \citep{Sanchis-Ojeda2013}, larger radial velocity signals \citep{Pepe2013}, and high transit probability \citep{Winn2011}. As a result, many have been observed and characterized, and they make up a disproportionately high fraction of the known rocky planets with precisely measured densities \citep[e.g.,][]{jackson2013, Sanchis-Ojeda+14,Adams+16,Adams+17,Dai+17,Dai+18,Winn+18,Malavolta+18,Frustagli+20,Xiu+20}. USPs are also often found with further-orbiting companion planets \citep[e.g,.][]{Sanchis-Ojeda+14, Adams+21} Over the past $10$ years, already over a hundred of detections of such planets and planet candidates have been made \citep[][]{Sanchis-Ojeda+14,2020ChA&A..44..283X,2017AJ....154..226D}, with over four hundred TESS Project Objects of Interest with sub-$1$ day periods \citep[][]{exofop}. 

 Observations suggest a similar formation pathway for USPs and their longer period multiple planets counterparts \citep{mulders2016, Winn2017}. However, from theoretical point of view, USP planets pose a challenge for many planet formation models \citep[e.g.,][]{jackson2013, armstrong2020}. USPs can have significant orbital misalignments in multiplanet systems potentially explicable by the stellar quadrupole moment $J_2$ \citep[e.g.,][]{Becker+20,Li+20,chen+22}. The existence of the observed radius gap from $1.5$-$2.0$ Earth radii \citep[e.g.,][]{Fulton+17, Petigura+20,Loyd+20} may be connected to USPs via the large stellar flux they receive, driving off their atmospheres and lowering their radii \citep[e.g.,][]{Frustagli+20,Malavolta+18}.
 
 

Orbiting so close to their host stars, USPs are uniquely affected by short-range forces that are less significant in spread-out systems like the Solar System. One such force arises from the non-sphericity and non homogeneity of the host star, captured in the multipole expansion of the star's gravitational potential and represented by the moments $J_n$. To leading order, the most impactful moment is the gravitational quadrupole moment $J_2$ (hereafter referred to simply as $J_2$). A typical Sun-like star's $J_2$ is primarily created by centrifugal forces arising from its spin. As a star ages and its spin slows due to magnetic breaking, its $J_2$ decreases as well--all else equal. Importantly, a star's $J_2$ can have a significant impact of the structure and evolution of close-orbiting planets \citep[e.g.,][]{Spalding+16,Becker+20,chen+22}. 

There are mainly two approaches adopted in the dynamical analysis of USPs and multi-planet systems. One is exploring the planet-planet interactions in multi-planet system \citep[often including short range forces such as tides or general relativity][]{Liu+15,Denham+19,wei+21,Faridani+22}. The other considers the effect of the aforementioned stellar quadrupole  \citep[e.g.,][]{Spalding+16,Becker+17,Becker+20,Li+20,chen+22}. Thus, a natural question is how USPs and short-period planets are dynamically affected by these short-range forces when they exist within multi-planet systems. 

Here we show that the planet-planet interactions in a multi-planet system combined with evolving stellar spin can affect the stability and detectability of (ultra-)short-period planets. The planet-planet interactions excite high eccentricities and inclinations for specific USPs separations via Laplace-Lagrange resonance. However, stars spin down via magnetic braking as they evolve \citep[e.g.,][]{Dobbs-Dixon+04}. As the star's spin down, it's quadrupole moment, $J_2$, decreases, thus, sweeping the resonance locations from high semimajor axes to low. In other words, a short-period planet located in a stable zone, may undergo eccentricity and inclination excitation that may push it beyond transiting limits, or even destabilize the system.  


The paper is structured as follows. In section \ref{sec:equations} we outline the mathematical formalism quantifying how $J_2$ and GR affect the orbital evolution of the system. In section \ref{sec:j2res} we show the impact of the secular resonances that arise in multi-planet configurations and how decreasing $J_2$ values over time from stellar magnetic breaking can shift and strengthen or weaken the effect of secular resonances on close-orbiting planets. In section \ref{sec:stab} we give a summary of how we treat (in)stability of short-period planets in our simulations. In section \ref{sec:nominal} as a case study, we run a Monte Carlo suite of simulations on a proof-of-concept system based on HD 106315 and show the effects of $J_2$ on the stability and detectability of a potential planet orbiting inward of the planets observed in that system.  In section \ref{sec:hd106315} we explore the observed system HD 106315 and mark where in that system an inner, Earth-mass planet could remain stable and transit. Finally, in section \ref{sec:disc} we summarize our conclusions.


\section{Key physical processes}\label{sec:equations} 
Motivated by current observations we adopt a multi-planet system with small mutual inclinations and eccentricities \citep[e.g.,][]{Weiss+18Peas}, orbiting a star with mass $M$.  The time derivatives of $e$, $I$, $\Omega$ and $\varpi$ on a planet indexed by $j$ from Laplace-Lagrange are given by \citep[e.g.,][]{Murray+00book}
\begin{equation}\label{eq:LL_orbital_elements}
\begin{aligned}
\dot{e}_j & =-\frac{1}{n_j a_j^2 e_j} \frac{\partial \mathcal{R}_j^{(\rm sec)}}{\partial \varpi_j}, & \dot{\varpi}_j=+\frac{1}{n_j a_j^2 e_j} \frac{\partial \mathcal{R}_j^{(\rm sec)}}{\partial e_j}, \\
\dot{I}_j & =-\frac{1}{n_j a_j^2 I_j} \frac{\partial \mathcal{R}_j^{(\rm sec)}}{\partial \Omega_j}, & \dot{\Omega}_j=+\frac{1}{n_j a_j^2 I_j} \frac{\partial \mathcal{R}_j^{(\rm sec)}}{\partial I_j} ,
\end{aligned}    
\end{equation}

where $a_j$, $e_j$, and $I_j$ represent the semimajor axis, eccentricity, and inclination of planet $j$ respectively. $\mathcal{R}^{(\rm sec)}$ represents the averaged, secular disturbing function, which, in the Laplace-Lagrange, is given by,
\begin{equation}
\begin{aligned}
\mathcal{R}_j^{(\rm sec)}=& n_j a_j^2\bigg[\frac{1}{2} A_{j j} e_j^2+\frac{1}{2} B_{j j} I_j^2  \\  
&+\sum_{k=1, k \neq j}^N A_{j k} e_j e_k \cos \left(\varpi_j-\varpi_k\right) \\ 
&+\sum_{k=1, k \neq j}^N B_{j k} I_j I_k \cos \left(\Omega_j-\Omega_k\right) \bigg] ,
\end{aligned}    
\end{equation}
where



\begin{equation}
    A_{jj} = \frac{n_j m_k}{4\pi \left( M+m_j\right)} \alpha_{jk} \overline{\alpha}_{jk} f_\psi  \ ,
    \label{eq:A_matrix}
\end{equation}
\begin{equation}
    A_{jk} = -\frac{n_j m_k}{4\pi \left( M+m_j\right)} \alpha_{jk} \overline{\alpha}_{jk} f_{2\psi}  \ ,
\end{equation}
\begin{equation}
    B_{jj} = -\frac{n_j m_k}{4\pi \left( M+m_j\right)} \alpha_{jk} \overline{\alpha}_{jk} f_\psi  \ ,
\end{equation}
\begin{equation}
    B_{jk} = \frac{n_j m_k}{4\pi \left( M+m_j\right)} \alpha_{jk} \overline{\alpha}_{jk} f_\psi  \ ,
\end{equation}
and,
\begin{equation}
    \alpha_{jk} = \min \left(\frac{a_j}{a_k},\frac{a_k}{a_j}  \right) \ ,
\end{equation}
\begin{equation}
    \overline{\alpha}_{jk} = \min \left(\frac{a_j}{a_k},1  \right) \ ,
\end{equation}
and 
\begin{equation}
    f_{\psi}=\int_{0}^{2 \pi} \frac{\cos \psi}{\left(1-2\left(\frac{a_{1}}{a_{2}}\right) \cos \psi+\left(\frac{a_{1}}{a_{2}}\right)^{2}\right)^{2}} \mathrm{d} \psi\ ,
\end{equation}
and
\begin{equation}
    f_{2 \psi}=\int_{0}^{2 \pi} \frac{\cos 2 \psi}{\left(1-2\left(\frac{a_{1}}{a_{2}}\right) \cos \psi+\left(\frac{a_{1}}{a_{2}}\right)^{2}\right)^{\frac{3}{2}}} \mathrm{d} \psi \ .
\end{equation}



The effect of General Relativity (GR) precession on the $\omega$ of a planet with semimajor axis $a$ and eccentricity $e$ is given by \citep[e.g.,][]{Naoz+12GR}
\begin{equation}\label{eq:GR_orbital_elements}
    \dot{\omega}_{GR} = \frac{3 (GM_\star)^{3/2}}{c^2 a^{5/2} (1-e^2)} \ ,
\end{equation}
where $c$ is the speed of light.

The effect of the stellar gravitational quadrupole moment's ($J_2$) effect on $\varpi$ and $\Omega$ on a planet with semimajor axis $a$ and mean motion $n$ is given by \citep[e.g.,][]{Murray+00book}

\begin{equation}\label{eq:j2_orbital_elements}
\begin{aligned}
&\dot{\varpi}_{J_2}=+n\left[\frac{3}{2} J_{2}\left(\frac{R_{\mathrm{\star}}}{a}\right)^{2}\right] , \\
&\dot{\Omega}_{J_2}=-n\left[\frac{3}{2} J_{2}\left(\frac{R_{\mathrm{\star}}}{a}\right)^{2}\right] .
\end{aligned}
\end{equation}

\section{Secular Resonances of Eccentricity and Inclination from $J_2$}\label{sec:j2res}

When two planets' precessions of $\varpi$ or $\Omega$ are similar, they enter a secular resonance in $e$ or $I$, respectively \citep[e.g.,][]{Murray+00book}. This causes them to exchange angular momentum at a significantly higher rate than when they were out of resonance. Two planets can typically only enter secular resonance in systems with $3$ or more planets, though \citet{Spalding+16} showed that under the influence of $J_2$ (or, through the same logic, GR), the two planets in a two-planet system can enter secular resonance if the inner planet has more angular momentum than the outer planet.

However, the effect of $J_2$ on secular resonances is not only significant in two-planet systems. In systems with $3$ or more planets, where the innermost planet has the least mass, there are critical semimajor axis where it is in secular resonance with one of its companion planets. The impact of $J_2$ on this is to shift the values of the critical semimajor axis. Because the host star's spin slows over time due to magnetic breaking, this causes the shifts of the critical semimajor axis over time. 

\begin{figure*}
    
    \includegraphics[width=0.961\linewidth]{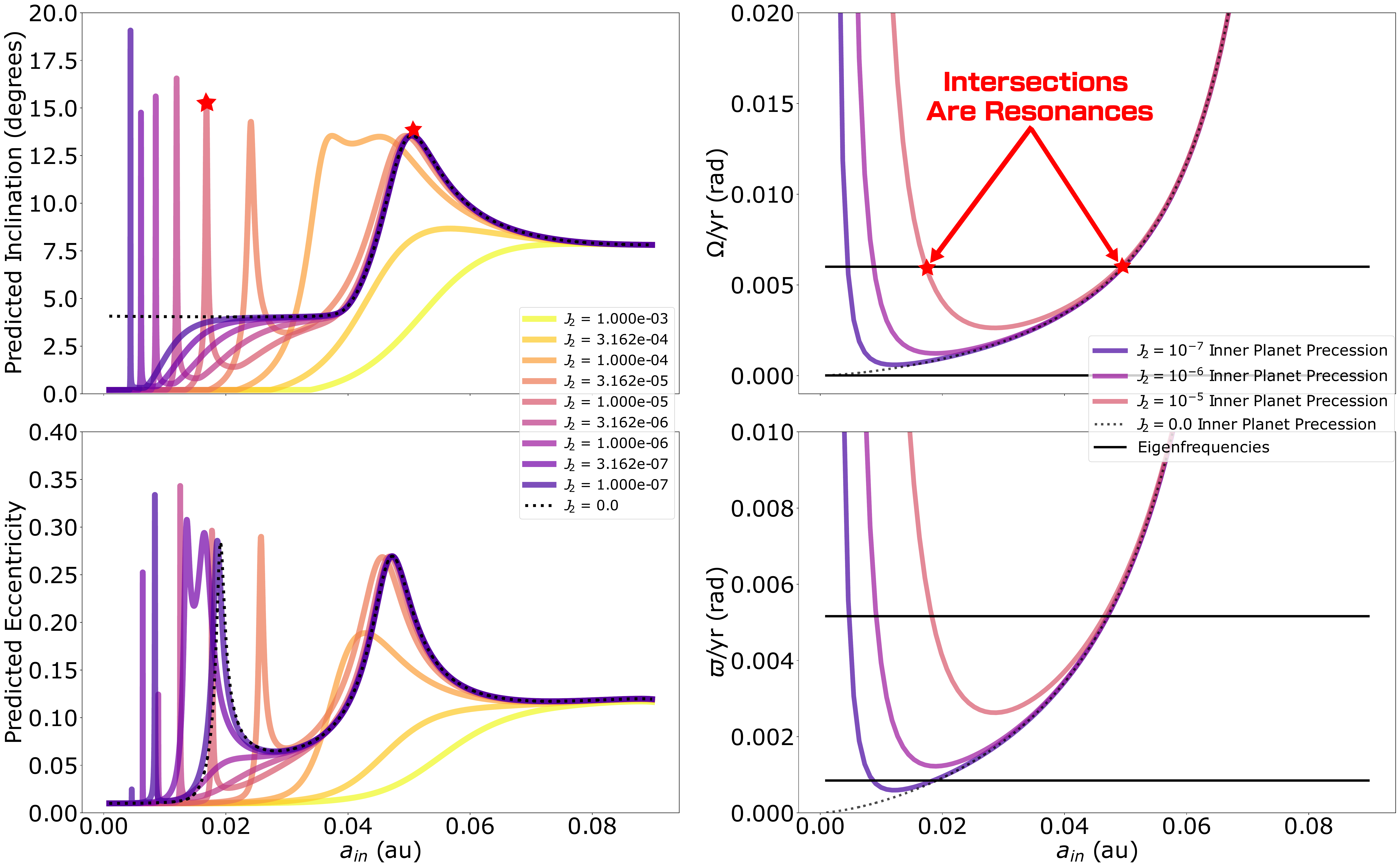}
    \caption{The left column shows the maximum eccentricity and maximum inclination of a hypothetical planet orbiting inward of our proof-of-concept system's two planets expected by Laplace-Lagrange for various values of $J_2$. At small semimajor axes ($<0.04~$au) there exists a value of $J_2$ such that the combined Laplace-Lagrange and $J_2$ effect produces a resonance at that semimajor axis. The orbital parameters of the system here are shown in Table \ref{tab:proofofconceptparams}, with the hypothetical planet having a mass of $1~M_\oplus$, eccentricity $e=0.05$, and inclination $i=1^\circ$. The right column shows, for several values of $J_2$ the impact of inner planet semimajor axis on the inner planet's precession of $\varpi$ (top) and $\Omega$ (bottom). Also shown are the eigenfrequencies of the secular matrices $B$ (top) and $A$ (bottom). When the inner planet's precession of $\varpi$ or $\Omega$ intersects with an eigenfrequency, that corresponds to a secular resonance. See for example, how this representation explains why the top left panel is simpler than the bottom left panel--there are fewer crossings with the other planets' eigenvalues at low $J_2$. The critical $a_{\rm in}$ values that are in resonance for $J_2=10^{-5}$ are marked in the top row to demonstrate the relationship between intersections in the right column and resonances in the left column. An animated version of the left column is available. The animation evolves this curve continuously from $J_2=10^{-3}$ to $10^{-7}$, clearly demonstrating the structure.} 
    \label{fig:j2 sweep}
\end{figure*}

This is shown in Figure \ref{fig:j2 sweep}, where, in the left column we plot the expected inner planet inclination and eccentricity from secular interactions as functions of inner planet semimajor axis for different values of $J_2$ for the system described in Table \ref{tab:proofofconceptparams}. These values are calculated analytically from the equations in Section \ref{sec:equations} with the effects of $J_2$ included in the $A$ and $B$ matrices \citep[see e.g., the construction in][]{Murray+00book}. In the right column we show the precession of the inner planet's longitude of ascending node ($\Omega$, top) and longitude of periapsis ($\varpi$, bottom), as well as the eigenfrequencies of the secular matrices B (top) and A (bottom) as a funcion of the inner planet's semimajor axis. When the inner planet's precession intersects an eigenfrequency for a particular values of $J_2$, this corresponds to a secular resonance, dramatically increasing the eccentricity or inclination excited by the outer planets on the inner planet. In the top row we mark the resonances and intersections for the curves where $J_2=10^{-5}$ as an example. Notably, there can only ever (in this system) be two intersections in the $\Omega$ precessions, corresponding to there only ever being up to two peaks in the inclination plots. However, the larger, lower eigenfrequency in the $\varpi$ plot means that there can be up to $4$ peaks in excited eccentricity for a particular $J_2$.

At different values of $J_2$ the peaks in these plots representing the secular resonances shift dramatically. This implies that if a star's $J_2$ significantly evolves over its lifetime, that at some point during that evolution, $J_2$ will have the appropriate value to send an inner planet into secular resonance for a large range of inner planet semimajor axes.

\section{Stability of Short-Period Planets}\label{sec:stab}

Short-period planets are subject to many dynamical effects, including GR,  planet-planet interactions, and precession due to stellar oblateness\footnote{Tidal effects such as precesssion, circularization and shrinking of the planet's orbit may also contribute to the orbital stability, but they are beyond the scope of this paper. We expect that tidal precession will be of the order of GR precession timescale and thus both will contribute similarly to the stability of the system. }All of these effects can impact the stability of a short-period planet. Secular resonances through planet interactions can cause rapid angular momentum exchange, driving up inner planets' eccentricities \citep[e.g.,][]{Faridani+22}. In multi-body systems with far-orbiting outer bodies, angular momentum exchange through the Eccentric Kozai-Lidov-like effect can even more efficiently destabilize an inner planet or body \citep[e.g.,][]{Naoz16}. These eccentricity excitations can be suppressed if GR precession (or other short range forces) takes place faster than the multi-body orbits, and the far away companion, induce precession
\citep[e.g.,][]{Denham+19, wei+21}. Sufficiently high values of $J_2$ can destabilize short-period planets with only one companion \citep[e.g.,][]{Spalding+16}, though in this work we will be focusing on short-period planets with 2 (or more) companion planets.

In this work, we characterize the stability of an inner planet over the course of its lifetime with an index, $\delta$, 
\begin{equation}
    \delta = {\rm max} \left(\frac{r_{{\rm peri},1}-r_{{\rm apo},{\rm in}}}{a_1-a_{\rm in}},0\right),
    \label{eq:delta}
\end{equation}
where $a_{\rm in}$ and $r_{{\rm apo},{\rm in}}$ represent the semimajor axis and apocenter distance of the inner planet, respectively, and $a_1$ and $r_{{\rm peri},1}$ represent the seimajor axis and the pericenter distance respectively of the closest-orbiting planet to the inner planet. Here, $\delta$ represents, over its whole lifetime, the closest an inner planet got to crossing orbits with its closest companion planet.
A $\delta$ of $0$ represents a potential-orbit crossing scenario, where the Laplace-Lagrange formalism no longer applies and the system is likely unstable.

Here, orbital crossing is treated as a catastrophic fate for the system, which may lead to the ejection of one or more planets (or changing their orbital energies by a factor of two).  However, orbital crossing does not yield an instantaneous destabilization of the system. In fact it was recently shown that orbits within each other's Roche limit, and even crossing orbits, have a timescale for destabilization, in some cases longer than the system's lifetime \citep{Zhang+23}\footnote{Non hierarchical systems, but perhaps not orbit crossing, do not instantaneously destabilize  \citep[e.g.,][]{Grishin+18,Mushkin+20,Bhaskar+21}. Thus, even low values of $\delta$ may take time to undergo dramatic changes for their orbits. }. However, here for simplicity, when running simulations, we stop the simulation when the orbits cross.

\begin{figure}
    \includegraphics[width=\linewidth]{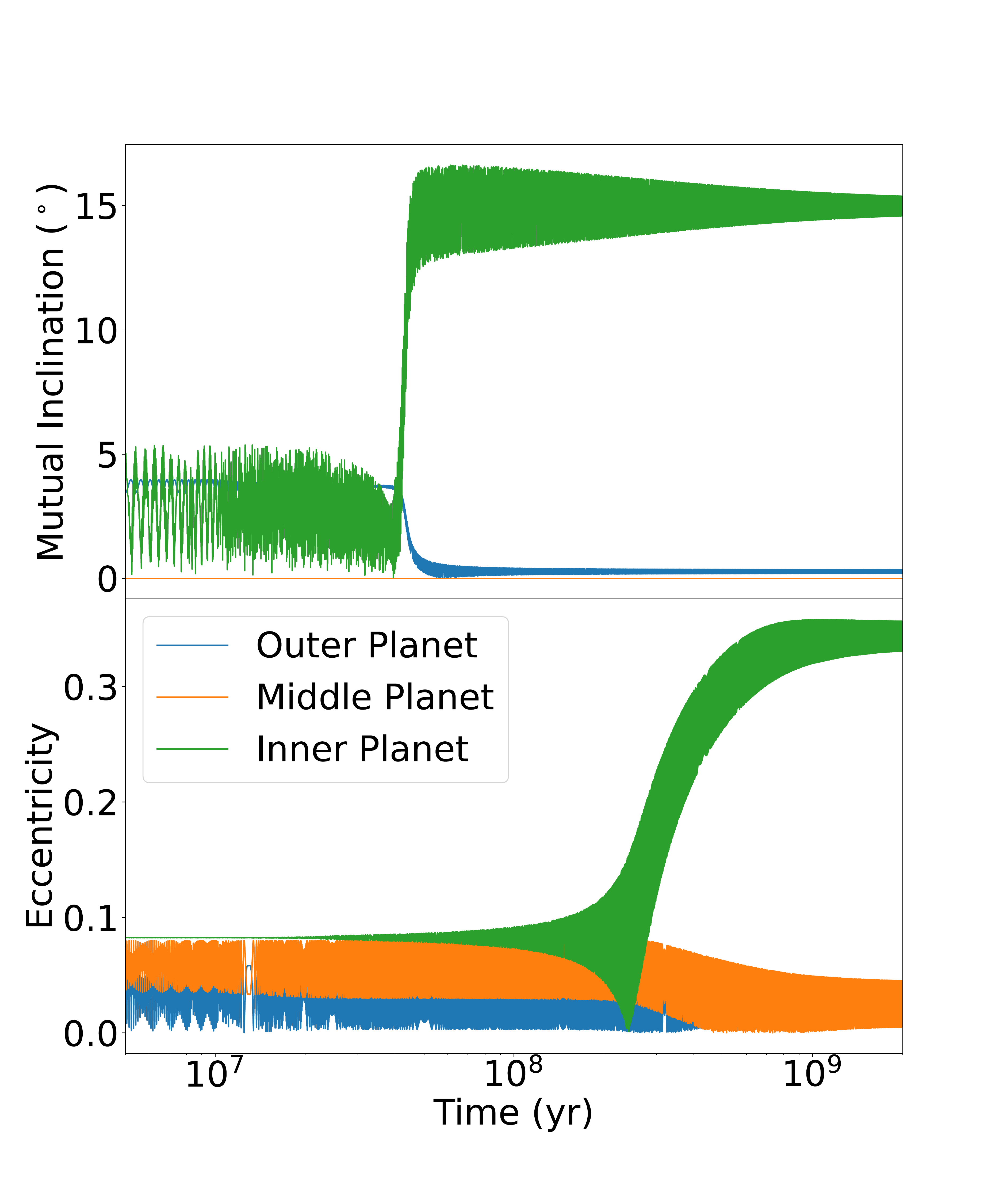}
    \caption{
    Eccentricity and mutual inclination with middle planet as a function of time for a run from the Monte Carlo described in Section \ref{monte}. An inner planet was inserted into the proof-of-concept system with initial conditions $a_{\rm in}=0.02061$, $e_{\rm in}=0.08247$, $i_{\rm in}=1.07865^\circ$,$\Omega_{\rm in}=154.6^\circ$. Representative of the effect of sweeping resonances is the swift transition in the inclination of the inner planet as a resonance passes through it at $\sim 40~$Myr. The rolling average of the inclination of the middle planet remains at $\sim 2.5^\circ$ over the the entire integration, with the magnitude of its inclination oscillations changing from $\pm 4^\circ$ to $\pm 0.5^\circ$ after the inner planet transitions to its higher mutual inclination state. The slower climb of the eccentricity evolution is caused by the fact that the inner planet's semimajor axis corresponds to a resonance when $J_2\sim 0$, meaning that as $J_2$ decreases over the whole integration, the predicted eccentricity slowly increases rather than suddenly transitioning. }
    \label{fig:time_evolution}
\end{figure}

\section{Proof of Concept System}\label{sec:nominal}
\subsection{System Description}
Here we consider a constructed system to demonstrate the impact of secular resonances under the influence of evolving $J_2$. This system is based on the real system HD 106315 \citep[see e.g.,][]{Barros+17}. HD 106315 has significant eccentricity in both its observed planets, and has observed stellar obliquities of $\sim 10^\circ$ for both planets \citep[][]{Zhou+18}. These properties make HD 106315 very interesting, though too dynamically hot to act as a representative example of exoplanet systems. To demonstrate the importance of $J_2$ on secular resonances for a wider variety of systems, we first explore a modified version of HD 106315 with much lower planetary eccentricities and stellar obliquities. HD 106315 was chosen as a base for this work because the spacing and moderate masses of its planets are ideal for showcasing the effect of evolving $J_2$ on secular resonances. 

HD 106315 is a $1.09\, M_\odot$ star with two observed planets detected via transits. HD 106315 b has semimajor axis, mass and radii $0.097\,$au ,$12.6\, M_\oplus$, and $2.44 R_\oplus$ respectively. HD 106315 c has semimajor axis, mass and radii $0.1536\,$au ,$15.2\, M_\oplus$, and $4.35 R_\oplus$ respectively. The two planets have measured mutual inclinations of $<5^\circ$ \citep[][]{Zhou+18}. As a better proof-of-concept, we prepare both planets with eccentricites of $0.05$. The parameters of our proof-of-concept system are summarized in Table \ref{tab:proofofconceptparams}. 
To demonstrate the effect of secular resonance evolution, we will introduce a third, Earth-mass planet into the system orbiting within the two observed planets. We will explore the stability and detectability of this introduced inner planet under the influence of evolving resonances.

\begin{figure*}[p]
    \includegraphics[width=\linewidth]{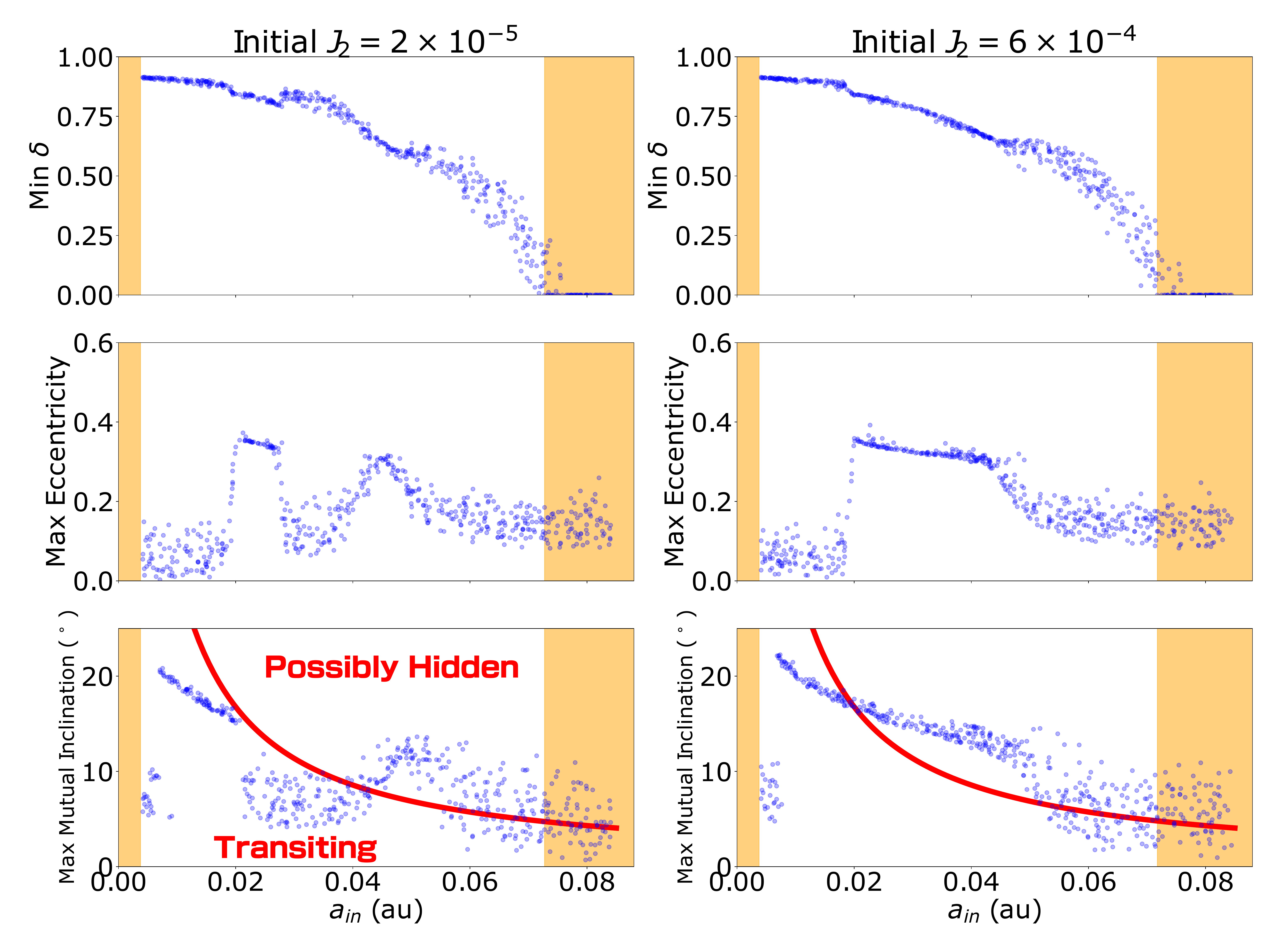}
    \caption{{\bf Proof of Concept System's inner planet.} Here we show the results of 500 systems, integrated up to 2 Gyr of our proof of concept system with an Earth-mass planet introduced inward of the two observed warm Neptunes. The left column represents initial $J_2=2\times 10^{-5}$ while the right column represents $J_2=6\times 10^{-4}$, these two values represent the bottom and top curves in Figure \ref{fig:j2_evolution}. 
    The top row indicates for each run the minimum value of $\delta$ (see Eq. \ref{eq:delta}) achieved over the entire run. The shaded regions represent regions of instability either from Roche limit crossing (left) or orbit-crossing indicated by $\delta = 0$ (right). The middle row indicates the maximum eccentricity achieved by the inner planet over each run. The bottom row indicates the maximum mutual inclination between the inner planet and the two warm Neptunes over the course of the run. The red curve indicates the maximum mutual inclination that the inner planet can have with its closest neighbor and still transit.
    }
    \label{fig:Earth6Plots}
\end{figure*}

\begin{table}
\tiny
\centering
\begin{tabular}{lccccc}
\hline \hline Object & Mass $\left(M_{\odot}\right)$ & $a(\mathrm{au})$ & $e$ & $i$ & $\Omega$ \\
\hline Star & $1.09$ & & & \\
Inner Planet & $1.0~M_\oplus$ & $0-0.085$ & $0-0.2$ & $0-7^\circ$ & $0-360^\circ$ \\
Planet b & $12.6~M_\oplus$ & $0.097$ & $0.05$ & $4^\circ$ & $0^\circ$ \\
Planet c & $15.2~M_\oplus$ & $0.1536$ & $0.05$ & $1^\circ$ & $0^\circ$ \\
\hline
\end{tabular}
\caption{Parameters of Proof-of-Concept System}
\label{tab:proofofconceptparams}
\end{table}

In Figure \ref{fig:time_evolution}, we show the evolution of the eccentricity and inclination of the planets of our proof-of-concept system with an additional, Earth-mass inner planet included inside Planet b's orbit. This evolution, and all other evolutions in this work unless stated otherwise are performed by a straightforward integration of the sums of Equations (\ref{eq:LL_orbital_elements}), (\ref{eq:GR_orbital_elements}), and (\ref{eq:j2_orbital_elements}). Allowing the value of $J_2$ in Equation (\ref{eq:j2_orbital_elements}) to vary allow us to straightforwardly model the evolution of $J_2$ as a function of time. The short-period planet has initial conditions of  $a_{\rm in}=0.02061$, $e_{\rm in}=0.08247$, $i_{\rm in}=1.07865^\circ$,$\Omega_{\rm in}=154.64^\circ$. Notable in these evolutions is the sudden transition of the inner planet's inclination from its starting value to higher values at approximately $40$ Myr. This transition is caused by the evolution of $J_2$ bringing the inner planet into a secular resonance with its companions. Thus, as $J_2$ evolves, the inner planet's inclination can reach its resonant value, depending on its semi-major axis, as shown in Figure \ref{fig:j2 sweep} (highlighted by the dashed-dot line). The relatively fast ``jump'' in inclination occurs because the ranges of $J_2$ where the inner planet is in inclination resonance is narrow, and the there are no very many intermediate $J_2$ values between non-resonant and resonant ranges. This means that the transition from non-resonant to resonant happens quickly as $J_2$ continuously evolves. Further, as $J_2$ continues to evolve, the system leaves the resonance, but the high inclination remains as a signature of past resonance. The growth in inner planet eccentricity shown in Figure \ref{fig:time_evolution}, however, arises because the inner planet's semimajor axis is very close to a resonance at $J_2=0$, this means that as $J_2$ evolves over the $2\,$Gyr evolution, the inner planet grows closer and closer to resonance continuously, causing a much slower growth in eccentricity.


\subsection{Monte Carlo Exploration}\label{monte}
Here we present the results of $1000$ simulations of our proof-of-concept system with an inner, earth-mass planet  at a semimajor axis drawn from a uniform distribution $a_{\rm in}\in [0.005,0.085]$,  inward of the two observed planets. 
. 
The eccentricity and inclination (in radians) of the inner planet are drawn from Rayleigh distributions with $\sigma=0.049$, consistent with the observed distribution of small planets in Kepler multiplanet systems \citep[see e.g.,][]{VanEylen+15}. The short-period planet's longitude of ascending node and argument of periapsis were drawn from uniform distributions in $[0,2\pi]$.
 We integrate the Laplace Lagrange equation of motion including an evolving $J_2$, due to magnetic braking spindown, see Section \ref{appendic:j2}. We also include general relativity precession. 
 
All systems are integrated to $2~$Gyrs or until orbit crossing occurs. In Figure \ref{fig:Earth6Plots} we shown the maximum inclination (bottom row), eccentricity (middle row), and the minimum value of $\delta$, (see Equation (\ref{eq:delta}), top row). The latter represents the stability of the system, where zero corresponds to orbital crossing. Because $J_2$ depends on so many stellar parameters (the Mass, Radius, Spin, and stellar Love Number, see Eq \ref{eq:j2_formula}), we adopt two widely-separated values ($2\times 10^{-5}$ and $6\times 10^{-4}$) for the initial $J_2$ in our simulations (see Figure \ref{fig:j2_evolution}) to capture the breadth of possible initial conditions.

The evolution of $J_2$ over the life of the star causes excitations in the eccentricity of the inner planet that change strength and location with time. Compared to a host star with unchanging $J_2$, this can reduce the size of the stable regions of parameter space for the inner planet. The critical role that an evolving $J_2$ plays in the dynamics of the system is highlighted in Figure \ref{fig:Earth6Plots}. For example, when comparing to Figure \ref{fig:j2 sweep}, individual curves in Figure \ref{fig:j2 sweep} yields well defined semimajor axes corresponding to secular resonance for the short-period planet's eccentricity and inclination. However, as $J_2$ evolves, these resonances sweep across a large semimajor axis range, yielding a wide plateau in the maximum eccentricity (see middle row in Figure \ref{fig:Earth6Plots}).

In the bottom row of Figure \ref{fig:Earth6Plots} , the mutual inclination between the short-period planet and its two companion planets is plotted. 
There are two distinct behaviors between the high and low initial $J_2$ cases, (left and right, column respectively). Namely for the low initial $J_2$ case there is a clear discontinuity and minimum in mutual inclination between $0.02-0.04~$au and for the high initial $J_2$ case, there is a smooth transition in the same region. This behaviour arises because at $J_2 = 10^{-4}$, a secular resonance in inclination for the inner planet exists at $a_{\rm in}=0.04~$au (see top panel of Figure \ref{fig:j2 sweep}). However, for the lower initial value of $J_2$ in the left column, those resonances never appear because the initial value of $J_2$ is lower than the values where these resonances appear. This means that in the left column, inner planets orbiting near $0.04~$au will never experience a secular resonance in inclination, and therefore have lower inclinations than their closer-orbiting counterparts. 

An immediate consequence from the evolution of the inclination is that the transit probability of a high inclination inner planet is low. In particular, we adopt the simple geometric condition for the critical transiting inclination $i_{\rm mut, tran} < \tan^{-1}(R_\star/a_{\rm in}$, where $R_\star$ is the radius of the star. This boundary is marked with a red curve. As can be seen, Both cases (low and high initial $J_2$) have the possibility to detect inner planets as close as $\sim 0.03$~au, with the low initial $J_2$ allowing for the possibility to detect some further out, thanks to the aforementioned minimum in mutual inclination. The discontinuity at $0.005$~au in the inclination arises from $J_2$ dominates over the dynamics, analogous to the GR dynamics, which result in a discontinuity in eccentricity at $0.02~$au.  

Significantly, the eccentricity excitation can lead to destabilizing the system up orbital crossing.  To estimate this, as mentioned above, we adopt the parameter $\delta$, Equation (\ref{eq:delta}), shown in the top panel.  Notably, for inner planet semimajor axies larger than $\sim 0.07$~au, regardless of inner planet initial orbital configuration or initial $J_2$, the system reach orbit crossing. However, even many of the inner planets that don't cross orbits have significantly elevated eccentricities, as high as $0.4$, which can pose challenges for the long-term stability of the system.

\begin{figure}
    \includegraphics[width=\linewidth]{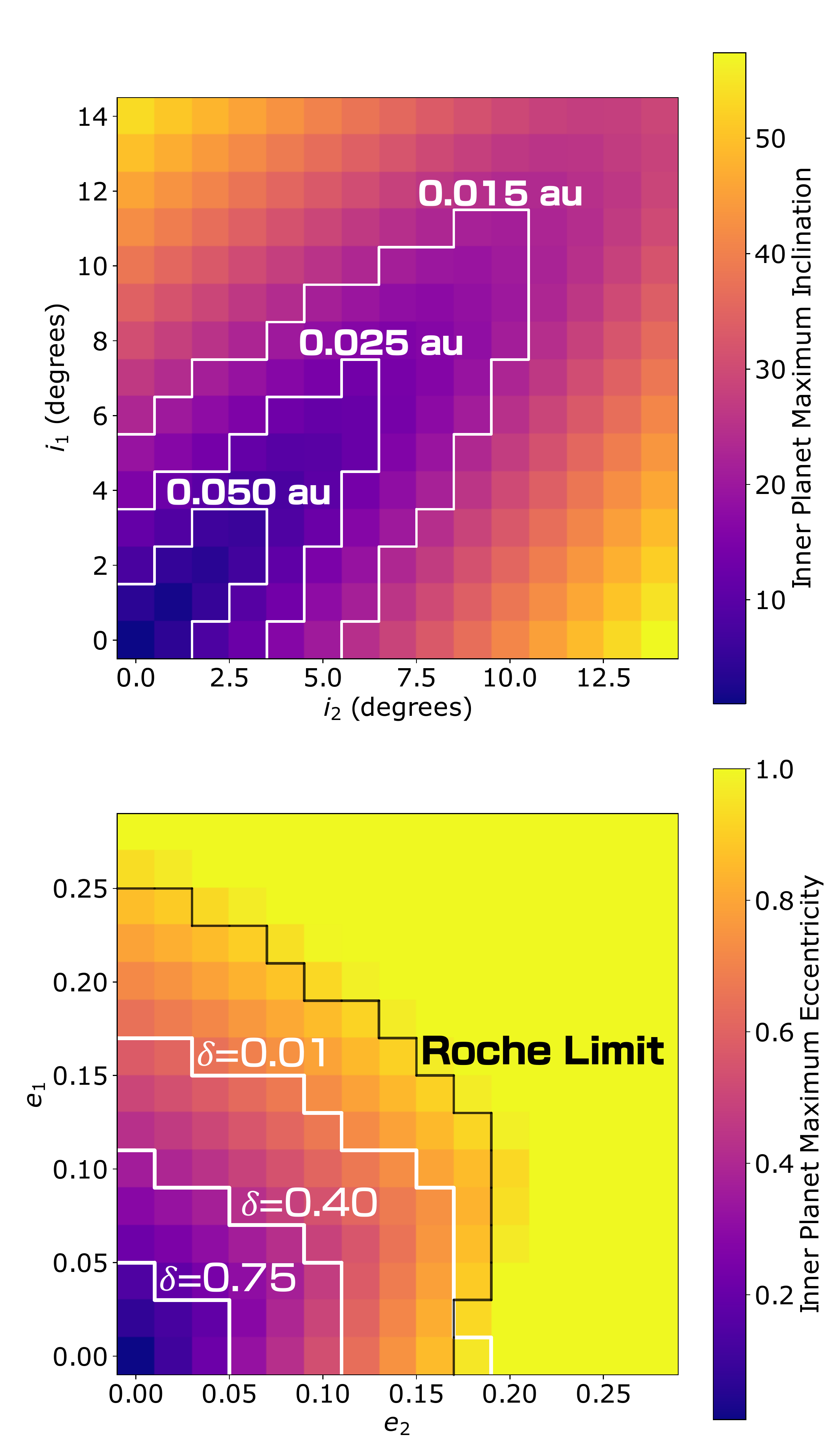}
    \caption{The maximum possible inclination and eccentricity achieved by an inner planet as a result of sweeping resonances given the inclinations/eccentricities of the outer planets. The reported maximum eccentricity and inclination values are found by calculating the predicted eccentricity and inclination excitation from a secular resonance. The inner planet's initial eccentricity and inclination are set to $0.01$ and $1^\circ$, respectively for these calculations. These maximum values are not significantly affected by $J_2$ evolution as $J_2$ serves primarily to move them rather than affect their magnitude (see Figure \ref{fig:j2 sweep}). Overplotted on the inclination panel are the contours within which the inner planet always transits not matter its longitude of ascending node, for the labeled semimajor axis. On the eccentricity panel are shown the contours within which $\delta$ is above the given value, for an inner planet with $a_{in}=0.046$. Additionally, the contour within which the inner planet does not cross the Roche Limit $R_L=R_p(2M_\star/M_p)^{1/3}$ for $a_{in}=0.046$.}
    \label{fig:colorplots}
\end{figure}

\subsection{Constraints on the Orbital Configuration of a Possible Inner Planet}
The eccentricity and inclination excitation from this channel can constrain the inner planet's orbital configuration, as well as its stability and detectability. 
As depicted in 
the top row of Figure \ref{fig:Earth6Plots}, for the representative example,  any such planets with semimajor axes greater than $0.07\, \text{au}$ undergo orbit crossing with their closest neighbor and are likely unstable. Those with semimajor axes between $0.02-0.04\, \text{au}$ become significantly eccentric ($e>0.3$) over the evolution of the system, indicating potential instability over the rest of the lifetime of the system from effects not captured by the first-order Laplace-Lagrange. 

Detectability via transit is also a concern for these type of planets. For example, in the simulations with high initial $J_2$, for planets with semimajor axes higher than $0.02\, \text{au}$, enough inclination is excited such that these planets may elude transit\footnote{Note that in Figure \ref{fig:Earth6Plots}, the transit boundary, depicted as the solid line in the Figure, represents the maximum inclination for which transiting can take place independently of $\Omega$. However, above the line, different values of $\Omega$ can yield a transiting system.  }, 
In the case with low initial stellar $J_2$, planets with semimajor axes higher than $0.045~$au may elude transit, but between $0.03-0.045~$au, a dip in maximum inclination exists. This dip arises because the inclination resonances 
only appear in the range $0.03-0.045~$au for $J_2$ values higher than $4\times 10^{-5}$ (see Figure  \ref{fig:j2_evolution} in Appendix \ref{appendic:j2}).

\subsection{The Effects of Initial Conditions}

As shown in Figure \ref{fig:Earth6Plots}, the main determinant in outcome for the inner planet with is its semimajor axis, as the spread around the trends in all 6 panels are relatively small. The exception to this are semimajor axes greater than $0.05~$au, where spreads are significant. This increased spread exists because no resonances exist at their semimajor axes for any values of $J_2$ (see Figure \ref{fig:j2 sweep}). This indicates that when a planet is affected by a sweeping resonance, that resonance determines its eccentricity and inclination excitation independent of any of its other orbital elements.

However, the effect of the initial eccentricities and inclinations of the outer two planets has great impact. In Figure \ref{fig:colorplots}, we show the maximum eccentricity and inclination of the inner planet over its evolution predicted by Laplace-Lagrange as functions of the initial inclinations and eccentricities of its two companions. Large eccentricities can be induced on the inner planet from moderate companion eccentricities, greatly impacting inner planet stability. Similarly, an inclination difference between the companions of two degrees can induce tens of degrees of misalignment on the inner planet. 

\section{HD 106315}\label{sec:hd106315}

\begin{figure*}[p]
    \includegraphics[width=\linewidth]{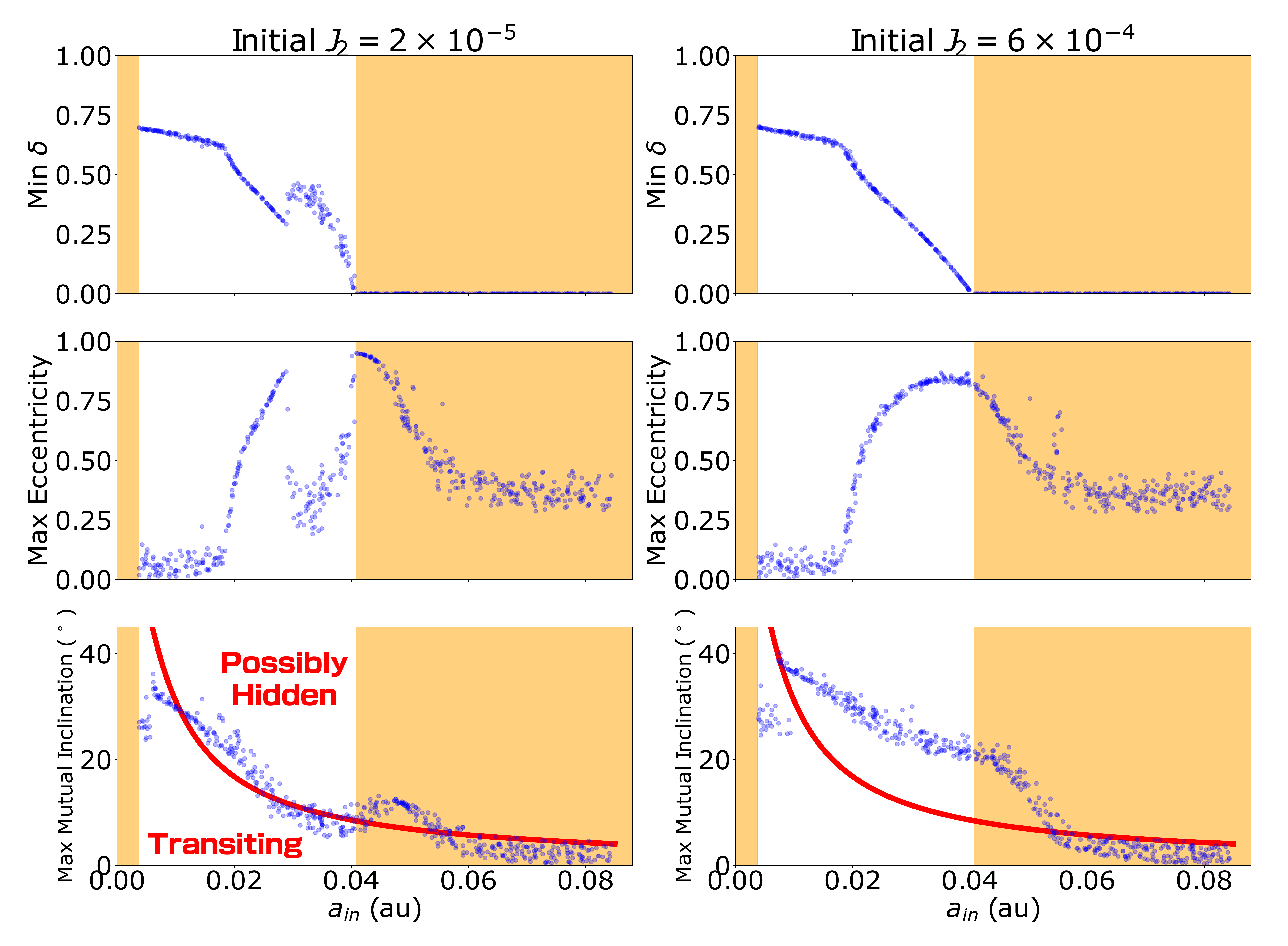}
    \caption{{\bf HD 106315's inner planet.} Here we show the results of 500 systems, integrated up to 2 Gyr of HD 106315 with an Earth-mass planet introduced inward of the two observed warm Neptunes. The left column represents initial $J_2=2\times 10^{-5}$ while the right column represents $J_2=6\times 10^{-4}$, these two values represent the bottom and top curves in Figure \ref{fig:j2_evolution}. 
    The top row indicates for each run the minimum value of $\delta$ (see Eq. \ref{eq:delta}) achieved over the entire run. The shaded regions represent regions of instability either from Roche limit crossing (left) or orbit-crossing indicated by $\delta = 0$ (right). The middle row indicates the maximum eccentricity achieved by the inner planet over each run. The bottom row indicates the maximum mutual inclination between the inner planet and the two warm Neptunes over the course of the run. The red curve indicates the maximum mutual inclination that the inner planet can have with its closest neighbor and still transit.}
    \label{fig:6plots106315Real}
\end{figure*}

Here we repeat the exercise of Section \ref{sec:nominal} except using the measured values of eccentricity and inclination for HD 106315.
\citet{Zhou+18} showed that the two warm Neptunes have low mutual inclinations ($< 5^\circ$, consistent with both choices of inclination described in Tables \ref{tab:proofofconceptparams} and \ref{tab:HD106315params}). Further, \citet{Zhou+18} also measured the obliquities of HD 106315 b and c, finding an obliquity of about $10^\circ$. We thus adopt these orbital parameters, and introduce an inner planet and ask what are the consequences of the aforementioned sweeping resonance on the possibility of the existence and orbital configuration of such a hypothetical planet. Particularly, in this exercise we are interested in analyzing the possibility of the system to hide an inner planet.

\begin{table}
\tiny
\centering
\begin{tabular}{lccccc}
\hline \hline Object & Mass  & $a(\mathrm{au})$ & $e$ & $i$ & $\Omega$ \\
\hline HD 106315 & $1.09~M_\odot$ & & & \\
Inner Planet & $1.0~M_\oplus$ & $0.005-0.085$ & $0-0.2$ & $9-16^\circ$ & $0^\circ$ \\
HD 106315 b & $12.6~M_\oplus$ & $0.097$ & $0.09$ & $14^\circ$ & $0^\circ$ \\
HD 106315 c & $15.2~M_\oplus$ & $0.1536$ & $0.22$ & $11^\circ$ & $0^\circ$ \\
\hline
\end{tabular}
\caption{Parameters used for the HD 106315 System \citep[][]{Barros+17}. Inclination values are consistent with \citet{Zhou+18}, who estimated the planets' mutual inclinations and measures their obliquity of $\sim 10^\circ$.}
\label{tab:HD106315params}
\end{table}

\subsection{Constraints on a Possible Inner Companion to HD 106315}

Figure \ref{fig:6plots106315Real} shows the Minimum $\delta$, and Maximum eccentricities and mutual inclinations for an inner planet introduced inwards of the two warm Neptunes of HD 106315. Notably, HD 106315 c's eccentricity of $0.22$ induces large eccentricities on the inner planet, as predicted by Figure \ref{fig:colorplots}. inner planets with semimajor axes higher than $0.045~$au have minimum $\delta = 0$, indicating that they cross orbits with the innermost warm Neptune. Moreover, even inner planets that don't cross orbits with the Neptunes have significantly elevated eccentricity unless they orbit at $<0.02~$au, where GR dominates, suppressing eccentricity excitation. 

In the case of low initial $J_2$, there is a region of (relatively) low eccentricity excitation from $0.03-0.04~$au, analagous to the same low-eccentricity region in the same place in Figure \ref{fig:Earth6Plots}. This region also has low mutual inclination with HD106315 b and c, with dots falling below the red curve, indicating that in this more stable region, an inner planet could survive and also transit with HD 106315 b and c. All in all, for the low initial $J_2$ case, stable and detectable (via transit) inner planets can exist in orbits $<0.01~$au and $0.03-0.04~$au. In the high initial $J_2$ case, all secular resonances in eccentricity for the inner planet play a role, eliminating the region of low-eccentricity present in the low initial $J_2$ case. For all orbits that have Min $\delta > 0$, only orbits with semimajor axis $<0.01~$au transit with HD 106315 b and c. 

Overall, the effect of sweeping resonances is to dramatically constrict the regime where a stable, detectable inner planet might live orbiting HD 106315. Inner planet orbits with semimajor axes greater than $0.04~$au cross with their closest neighbor, becoming unstable. For those that don't cross orbits, their detectability and stability are heavily affected by the initial $J_2$ of HD 106315 after the evaporation of the disk. If that initial $J_2$ is low enough, the secular resonances that would destabilize and incline planets in the $0.03-0.04~$au range do not form, making that a region where an undetected planet could conceivably survive. There is also the possibility of a stable and detectable orbiting at $<0.015~$au, however, at the time of writing, only two planets orbiting that close to a host star of mass $>1~M_\odot$ have ever been detected (PSR J1719-1438 b, and PSR J2322-2650 b \citep[][]{Bailes+11,Spiewak+18}), both orbiting pulsars (though PSR J1719-1438 b may be a carbon white dwarf), indicating that finding a short-period planet orbiting there is unlikely.

\section{Discussion}\label{sec:disc}

Compact multiplanet systems are common in our galaxy \citep[e.g.,][]{Howard+10,Howard+12,Lissauer+11,Winn+15}. Further, \citet{Weiss+18multis} suggested that planets in a system have similar sizes, called peas-in-a-pods \citep[but see, e.g.,][for possible selection biases]{Zhu20,Murchikova+20}. In fact, it was suggested that USPs are likely an extension of the population of hot Earth-sized planets \citep{mulders2016, Winn2017} to short periods. Thus, to develop a comprehensive understandings of USPs' dynamics it is imperative to investigate the combined effects of planet-planet interactions with evolving stellar quadrupole.

We demonstrated that the time-evolution of $J_2$ due to the stellar spin down can affect secular resonances in systems with at least $3$ planets\footnote{Note, that unlike, \citet{Spalding+16}, who considered $2$ planet systems, here we highlight the importance of the angular momentum exchange in a $\geq 3$ planet system, which forms a more complex resonant pattern. 
}. The quadrupole moment $J_2$'s evolution drives planets into and out of secular resonance as it evolves (illustrated in Figure \ref{fig:j2 sweep}). In systems like the ones we explored, where the innermost planet is significantly less massive than its companions, the effect of these sweeping resonances is to transition this inner planet to higher eccentricities and mutual inclinations over a relatively short period of time. This demonstrates the critical interplay between planet-planet interactions and star-planet interactions in determining the lifetime evolution of short-period planets.

We note that tidal effects become important over gigayear timescales. They were not included here as the timescales of tidal precession, circularization, and semimajor axis shrinkage are much slower than the GR, Laplace-Lagrange, and $J_2$ timescales. This means that tidal effects do not impede the effects of evolving $J_2$ to transition inner planets to higher eccentricities and misalign them from their companions. However, when considering the gigayears of evolution after secular resonances have affected the inner planet, tidal effects become significant, bringing the planet to shorter orbits and stabilizing it via circularization.

We showed the impact of sweeping secular resonances arising from evolving $J_2$ on a 3-body system with initially low mutual inclinations and eccentricities\footnote{
The impact of $J_2$ on two-planet systems has been well-demonstrated \citep[e.g.,][with continuous $J_2$ evolution in the latter]{Spalding+16,chen+22}, and has been explored in systems with more planets at several fixed $J_2$ values \citep[e.g.,][]{Li+20, Becker+20}, however we demonstrate with this system how in systems with 3 or more planets, the continuous evolution of $J_2$ is critical in understanding their evolution.} We began with a pedagogical proof of concept, highlighting the effect of sweeping resonances. For example, for a given semimajor axis for the innermost planet we depicted in Figure \ref{fig:time_evolution} the sharp eccentricity (inclination) transition as the planet enters its resonance, leaving a long lasting signature on the planet's eccentricity (inclination). 

These sweeping resonances can have large consequences on the planet's stability and transiting probability. This effect is demonstrated in Figure \ref{fig:Earth6Plots}, where we varied the innermost planet's semimajor axis neighbouring two outer planets.   
The destabilization can even occur when its period ratio with its closest neighbor is low. Moreover, the inner planet can potentially escape transit over a large semimajor axis range (see bottom panels of Figure \ref{fig:Earth6Plots}). This highlight the impact of sweeping resonances on the transiting probability. In particular, the other two planets orbital architecture does not vary significantly, remaining with their initial low inclinations and eccentricities, but the inner most planet, can evade transit. 
Using the aforementioned framework we 
investigated the stability and detectability of the observed system HD 106315. This system has two observed, transiting planets with an $\sim 10^\circ$ stellar obliquity, low mutual inclinations, and its outer planet has a moderate eccentricity of $\sim 0.22$ \citep[][]{Barros+17}. While there are no observed transits for an interior planet, we ask if the sweeping resonances can allow for the existence of a interior non-transiting planet. 

We find that in HD 106315, a potential Earth-mass inner planet is unstable at any semimajor axes greater than $0.04\,$au, and has high mutual inclinations with its transiting companions at semimajor axes inward of $0.025\,$au (see Figure \ref{fig:6plots106315Real}). This result depends on the initial $J_2$ of the star, which is largely unknown \citep[e.g.,][]{Baraffe+15,chen+22}. Specifically, a lower initial $J_2$ will yield only moderate mutual inclination, but similar stability regimes (see left column in Figure \ref{fig:6plots106315Real}).   
This is consistent with findings that transiting USPs and short-period planets often have companions \citep[e.g.,][]{Sanchis-Ojeda+14, Adams+21}.

Any approximately Earth-radius candidate transit detections in the HD 106315 system orbiting past $0.04\,$au could be reasonably discarded as unstable, and candidates orbiting at radii less than $0.025\,$au might be suspect due to their likely high mutual inclination with the two transiting planets. This result depends on the initial $J_2$ of the host star, with larger initial $J_2$ values reducing the size of the low-eccentricity region near $0.03\,$au and increasing the expected mutual inclination between the inner planet and its transiting companions.

\section*{Acknowledgements}
The authors thank Juliette Becker for useful discussions. S.N. thanks Howard and Astrid Preston for their generous support. G.L. is grateful for the support by NASA 80NSSC20K0641 and 80NSSC20K0522. 
This research has made use of the NASA Exoplanet Archive, which is operated by the California Institute of Technology, under contract with the National Aeronautics and Space Administration under the Exoplanet Exploration Program.


\begin{appendix}

\section{ $J_2$ Evolution}\label{appendic:j2}

\begin{figure}[h]
    \centering
    \includegraphics[width=0.65\linewidth]{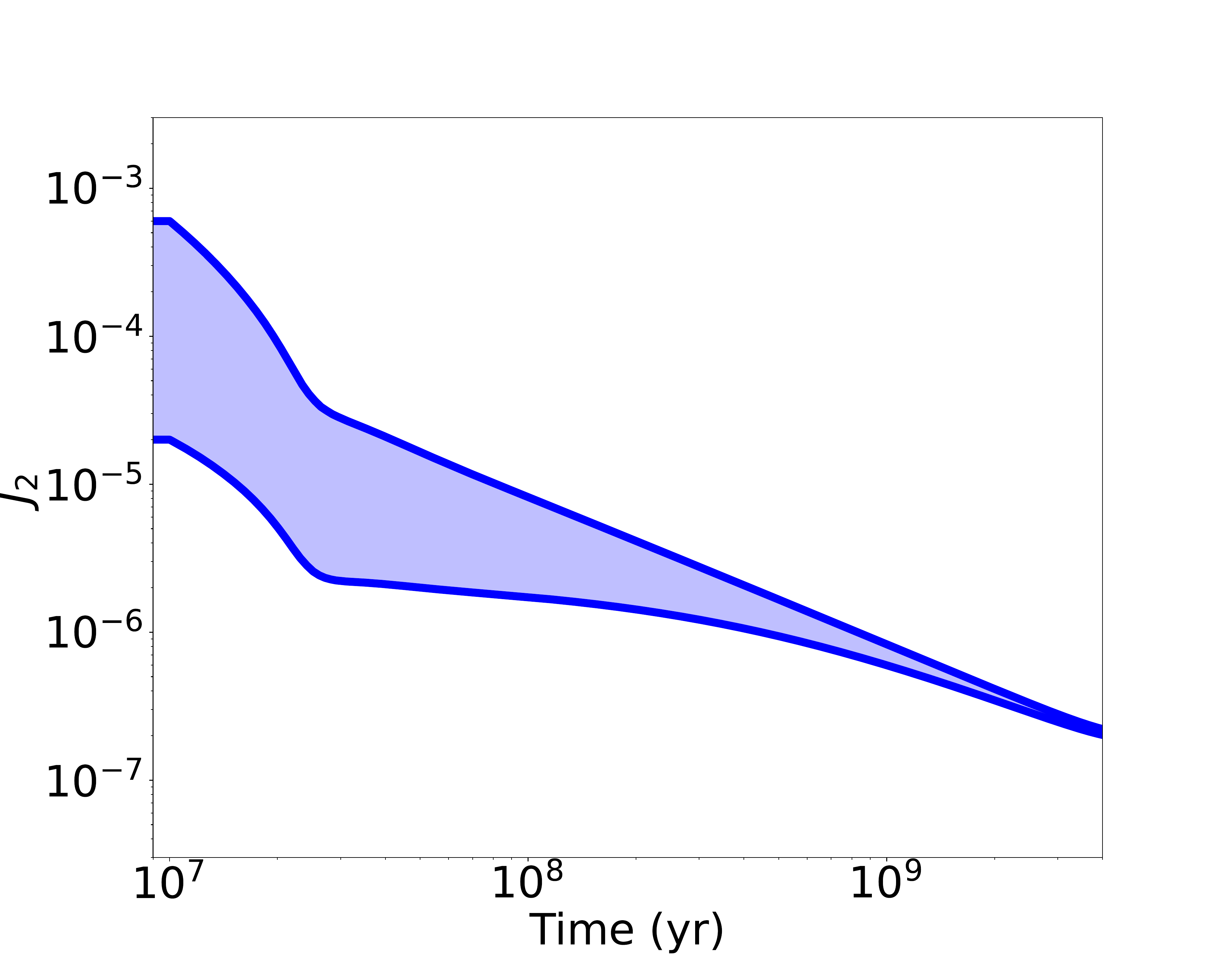}
    \caption{A range of possible $J_2$ evolutions for HD 106315 after the evaporation of the disk as a function of time. 
    }
    \label{fig:j2_evolution}
\end{figure}

To calculate the time evolution of the star's $J_2$, we use the relation that \citep[e.g.,][]{Spalding+16},
\begin{equation}
    J_2 = \frac{1}{3} k_2 \left(\frac{\Omega}{\Omega_b}\right)^2,
    \label{eq:j2_formula}
\end{equation}
where $k_2$ is the stellar Love number, $\Omega$ is the stellar spin, and $\Omega_b$ is the stellar breakup spin, $\Omega_b = \sqrt{GM/R^3}$.
The evolution of the Love number was calculated using a MESA stellar model \citep[][]{Paxton2011, Paxton2013, Paxton2015, Paxton2018, Paxton2019, Jermyn2023}, and the stellar spin evolution using the relation \citep[e.g.,][]{Dobbs-Dixon+04},
\begin{equation}
    \dot{\Omega} = -\alpha \Omega^3,
\end{equation}
with $\alpha = 1.5\times10^{-14}$. Because at late times ($>1~$Gyr), differences in initial $J_2$ have little impact on the final $J_2$, we chose a wide spread in our two initial $J_2$ values, $2\times 10^{-5}$ and $6\times 10^{-4}$. This choice demonstrates the impact of different initial $J_2$ values at early times.

\section{Comparison of Secular Code with N-body}\label{appendic:Nbody}

\begin{figure}[h!]
\centering
    \includegraphics[width=0.5\linewidth]{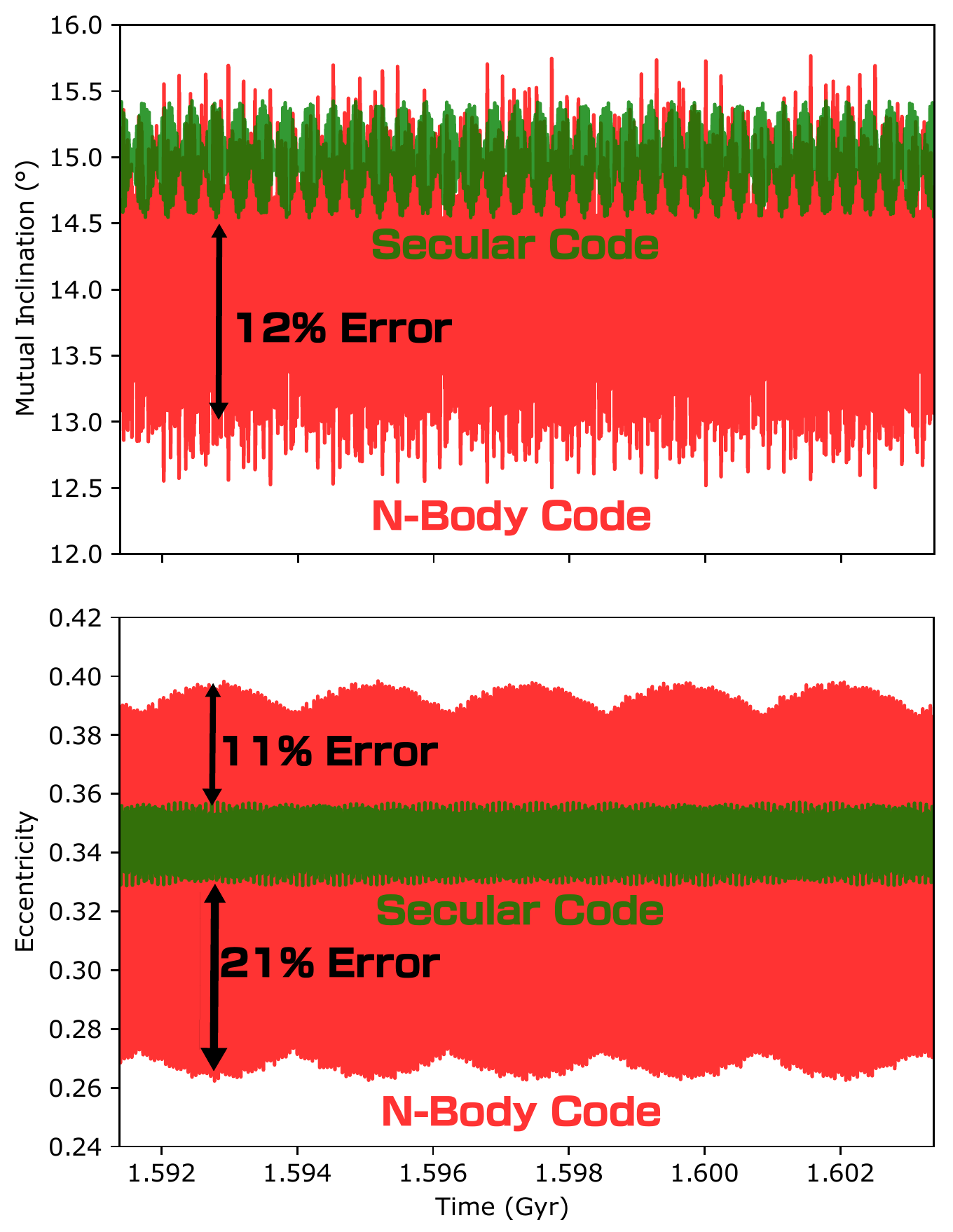}
    \caption{
    Comparison of the inner planet's eccentricity and mutual inclination with the middle planet when calculated with the secular code used throughout this work (green) and the N-body integrator HNBody (red). The green curve is a subsection of the evolution of the inner planet in Figure \ref{fig:time_evolution}, and the red curve was initialized with the orbital elements and $J_2$ value of all planets $1.591\,$Gyr into the evolution shown in Figure \ref{fig:time_evolution}.
    }
    \label{fig:sec-nbody-comparison}
\end{figure}

In Figure \ref{fig:time_evolution}, we show the results of our secular evolution. In Figure \ref{fig:sec-nbody-comparison}, we compare these results with the results from the N-body integrator {\tt HNBody}. The N-body simulations were prepared with initial conditions taken from the secular integration. We find that the N-body agrees well with the secular code's calculation of the maximum and mutual inclination achieved. However, the N-body shows that the magnitude of oscillations in eccentricity and inclination are significantly larger than predicted by the secular code, ranging from $\sim 10-20\%$ error, bringing eccentricity and inclination to lower minima. Additionally, {\tt HNBody} predicts a higher maximum eccentricity of $0.4$, a $11\%$ deviation from the secular code.

The impact of the differences between the N-body and secular code to the conclusions of this work is relatively minor. The lower minima in eccentricity oscillations does not significantly stabilize the system, as maximum eccentricity has much more impact on driving close encounters. The enhanced maximum eccentricity, however is not expected to significantly drive more collisions, as the $\delta$ values for similar ranges of eccentricity are probed in Figures \ref{fig:Earth6Plots} and \ref{fig:6plots106315Real}, and minimal differences in eccentricity are achieved. For inclination however, the differences might have more impact--affecting whether or not the inner planet transits with its companions. However, from the bottom row of Figure \ref{fig:Earth6Plots} and the top panel of Figure \ref{fig:colorplots}, a mutual inclination of $16^\circ$ is not enough to remove the inner planet from co-transiting geometry, indicating the the differences between the N-body and secular code are not qualitatively significant in this case.


\end{appendix}



\bibliographystyle{aasjournal}
\bibliography{kozai, exo} 






\bibliographystyle{aasjournal}

\end{document}